\documentstyle[12pt]{article}
\begin{document}

\author{Rui Carvalho \\
{\it Laborat\'orio de Mecatr\'onica, DEEC - IST,}\\
{\it \ Av. Rovisco Pais, 1096 Lisboa Codex, Portugal} \and R. Vilela Mendes%
\thanks{%
Corresponding author, e-mail: vilela@alf4.cii.fc.ul.pt} \\
{\it Grupo de F\'\i sica-Matem\'atica}\\
{\it \ Complexo Interdisciplinar, Univ. de Lisboa}\\
{\it \ Av. Gama Pinto,2 - 1699 Lisboa Codex, Portugal} \and Jo\~ao Seixas%
\thanks{%
On leave from Dept. F\'\i sica, Instituto Superior T\'ecnico, Av. Rovisco
Pais, 1096 Lisboa Codex, Portugal} \\
{\it Theory Division, CERN}\\
{\it \ CH 1211 Geneva 23, Switzerland}}
\title{Feigenbaum networks}
\date{}
\maketitle

\begin{abstract}
We study dynamical systems composed of a set of coupled quadratic maps
which, if uncoupled, would be on the Feigenbaum accumulation point.

For two units we prove the existence of an infinite number of sinks for an
open set of coupling parameters. In the limit of many units a mean field
analysis also implies the stabilization in periodic orbits of, at least, a
subset of the coupled units.

Possible applications in the fields of control of chaos, signal processing
through complex dynamics and as models of self-organization, are discussed.
\end{abstract}

\section{Introduction}

In this paper we study dynamical systems composed of a set of coupled
quadratic maps 
\begin{equation}
x_{i}(t+1)=1-\mu _{*}\{\sum_{j}W_{ij}x_{j}(t)\}^{2}  \label{1.1}
\end{equation}
with $x\in [-1,1]$, $\sum_{j}W_{ij}=1$ $\forall i$ and $W_{ij}>0$ $\forall
i,j$ and $\mu _{*}=1.401155...$. The value chosen for $\mu _{*}$ implies
that, in the uncoupled limit ($W_{ii}=1,W_{ij}=0$ $i\neq j$), each unit
transforms as a one-dimensional quadratic map in the accumulation point of
the Feigenbaum period-doubling bifurcation cascade. The coupling constants $%
W_{ij}$ resembling the synaptic connections of a neural network, we call
this system a Feigenbaum network. The system (\ref{1.1}) has a rich
dynamical behavior and coupled systems of this type have been used in the
past to model the dynamics of interacting populations\cite{Gyllenberg} \cite
{Hastings}.

The quadratic map at the Feigenbaum accumulation point is not in the class
of chaotic systems (in the sense of having positive Lyapunov exponents)
however, it shares with them the property of having an infinite number of
unstable periodic orbits. Therefore, before the interaction sets in, each
elementary map possesses an infinite diversity of potential dynamical
behaviors. As we will show later, the interaction between the individual
units is able to selectively stabilize some of the previously unstable
periodic orbits. The selection of the periodic orbits that are stabilized
depends both on the initial conditions and on the intensity of the
interaction coefficients $W_{ij}$. As a result Feigenbaum networks appear as
systems with potential applications in the fields of control of chaos,
information processing and as models of self-organization.

Control of chaos or of the transition to chaos has been, in recent years, a
very active field (see for example Ref.\cite{Shinbrot} and references
therein). Several methods were developed to control the unstable periodic
orbits that are embedded within a chaotic attractor. Having a way to select
and stabilize at will these orbits we would have a device with infinite
storage capacity (or infinite pattern discrimination capacity). However, an
even better control might be achieved if, instead of an infinite number of
unstable periodic orbits, the system possesses an infinite number of
periodic attractors. The basins of attraction would evidently be small but
the situation is in principle more favorable because the control need not be
as sharp as before. As long as the system is kept in a neighborhood of an
attractor the uncontrolled dynamics itself stabilizes the orbit.

The creation of systems with infinitely many sinks near an homoclinic
tangency was discovered by Newhouse\cite{Newhouse} and later studied by
several other authors\cite{Robinson} \cite{Gambaudo} \cite{Tedeschini} \cite
{Wang} \cite{Nusse}. In the Newhouse phenomenon infinitely many attractors
may coexist but only for special parameter values, namely for a residual
subset of an interval. Another system, different from the Newhouse
phenomena, which also displays many coexisting periodic attractors is a
rotor map with a small amount of dissipation\cite{Feudel}.

In Sect. 2 we prove that with only two units and symmetrical couplings one
obtains a system which has an infinite number of sinks for an open set of
coupling parameters. The mechanism by which an infinite number of unstable
periodic orbits is stabilized by the coupling of two units depends strongly
on the properties of the maps at the Feigenbaum accumulation point. A
similar mechanism would not operate, for example, in the tent map. The
mechanism by which stable periodic orbits might be obtained in piecewise
linear maps is quite different, as discussed in the last Section of the
paper.

Then, in Sect.3, we analyze the behavior of a Feigenbaum network in the
limit of a very large number of units. Using a mean field analysis we show
how the interaction between the units generates distinct periodic orbit
patterns throughout the network. The framework and motivation for studying
chaotic (or chaotic-like) networks for signal processing and pattern
recognition is then discussed, a simple example being presented in the
Feigenbaum network setting. Finally, in Sect.4, we add a few conclusions and
discuss some other systems, with similar behavior, which might be useful for
practical designing purposes.

\section{A simple system with an infinite number of sinks}

Here we consider two units with symmetrical positive couplings ($%
W_{12}=W_{21}=c>0$)

\begin{equation}  \label{2.1}
\begin{array}{c}
x_1(t+1)=1-\mu _{*}\left( (1-c)x_1(t)+cx_2(t)\right) ^2 \\ 
x_2(t+1)=1-\mu _{*}\left( cx_1(t)+(1-c)x_2(t)\right) ^2
\end{array}
\end{equation}

The mechanism leading to the emergence of periodic attractors from a system
that, without coupling, has no stable finite-period orbits is the permanence
of the unstabilized orbits in a flip bifurcation and the contraction effect
introduced by the coupling. The structure of the basins of attraction is
also understood from the same mechanism. The result is:

{\bf Theorem}: {\it For sufficiently small }$c${\it \ there is an }$N${\it \
such that the system (\ref{2.1}) has stable periodic orbits of all periods }$%
2^n${\it \ for }$n>N${\it .}

{\bf Proof}:

The bifurcations leading to the Feigenbaum accumulation point at $\mu _{*}$
are flip bifurcations. This means that, after each bifurcation, the orbit
that looses stability remains as an unstable periodic orbit. Therefore, (for 
$c=0$) at $\mu =\mu _{*}$ the system (\ref{2.1}) has an infinite number of
unstable periodic orbits of all periods $p=2^n$.

The proof has two basic steps. First we need to prove that, for sufficiently
small $c\neq 0$, these periodic orbits still exist in the system (\ref{2.1}%
). Second, that for any such $c$, there is an $N$ such that there is at
least one stable orbit for all periods $p=2^{n}$ with $n>N$. For both steps
an important role is played by the instability factor, given by $%
f^{(p)^{^{\prime }}}(x_{p})$ at the fixed points $x_{p}$ of the $p-$iterated
map. Therefore we will first find the behavior of this derivative for the
periodic unstable orbits at $\mu =\mu _{*}$ and $c=0.$

Iterating once the Feigenbaum-Cvitanovic functional equation 
\begin{equation}
-\frac{1}{\lambda }f\circ f(-\lambda x)=f(x)  \label{2.2}
\end{equation}
and taking derivatives one concludes 
\begin{equation}
f^{(4)^{\prime }}(x)=f^{^{\prime }}\left( \frac{x}{\lambda ^{2}}\right)
\label{2.3}
\end{equation}
Because $\lambda =0.3995...$ is the scaling factor that controls distances,
if, for example, $x^{*}$ is the fixed point of $f^{(p)}$ closest to zero,
the corresponding fixed point for $f^{(4p)}$ will be at $x^{*}\lambda ^{2}$.
Eq.(\ref{2.3}) means that for the scaling function $f$, solution of the
functional equation, the instability factor for all unstable periodic orbits
is the same at $\mu =\mu _{*}$. The logistic function is not a solution of
the functional equation, however it shares the scaling properties of this
solution at $\mu =\mu _{*}$. Therefore the instability factor $\left(
f^{(p)}\right) ^{^{\prime }}(x^{*})$ of the orbits at $\mu _{*}$ will also
converge to a fixed value. Numerically one finds this value to be around $%
-1.6$. The exact value is not important however, because what matters for
our result is that it is a non-zero uniformly bounded value for all orbits.
We may now proceed to the proof of the theorem.

{\it First step}: Permanence of the periodic orbits for small $c$

Let $x_{p}^{*}\in [-1,1]\times [-1,1]$ be, for example, the coordinate of
the $p-$periodic orbit closest to zero. The sequence $\left\{ x^{*}\right\}
=\left\{ x_{p}^{*}:p=2,4,8,...\right\} $ is an element of a $\ell _{\infty }$
Banach space (sup norm). The collection of fixed point equations 
\[
f_{\mu ^{*}}^{(p)}(x_{p}^{*},c)-x_{p}^{*}=0 
\]
defines a $C^{\infty }-$mapping $F(x^{*},c)$ from $\ell _{\infty }\times
R\rightarrow \ell _{\infty }$ . Because $\left( f^{(p)}\right) ^{^{\prime
}}(x^{*})$ at $c=0$ is negative and bounded for all $p$, the derivative $%
D_{1}F$ of the mapping in the first argument is invertible. Therefore, by
the implicit function theorem for Banach spaces, there is a $c^{*}$ such
that for $c<c^{*}$ the function $x^{*}(c):R\rightarrow \ell _{\infty }$ is
defined, that is, there are $p-$periodic orbits for all periods $p=2^{n}$.
The derivative of this function is defined componentwise (in $p$) by 
\begin{equation}
\frac{\partial x^{*}}{\partial c}=-\left( \frac{Df_{\mu _{*},c}^{(p)}}{Dx}%
-1\right) ^{-1}\frac{\partial f_{\mu _{*},c}^{(p)}}{\partial c}  \label{4}
\end{equation}

For the uncoupled case ($c=0$) the instability factor $\left( f^{(p)}\right)
^{^{\prime }}(x^{*})$ for each mapping is the product $\left( -2\mu
_{*}\right) ^{p}\prod_{k=1}^{p}x(k)$ over the orbit coordinates. For $%
c<c^{*} $, the orbit structure being preserved, their projections on the
axis are continuous deformations of the $c=0$ case which will preserve the
geometric relations of the Feigenbaum accumulation point. Hence the same
products for the projected coordinates suffer changes of order $a(c)p\lambda
^{p}$ and remain bounded.

{\it Second step}: Stabilization of at least one orbit for all periods $%
p=2^n $ with $n>N(c)$

The stability of the periodic orbits is controlled by the eigenvalues of the
Jacobian $J_{p}=\frac{Df_{\mu _{*},c}^{(p)}}{Dx}$ in the fixed point of $%
f_{\mu _{*},c}^{(p)}$. The map (\ref{2.1}) is a composition of two maps $%
f_{1}\circ f_{2}$%
\[
f_{1}:\left( 
\begin{array}{c}
x_{1} \\ 
x_{2}
\end{array}
\right) \rightarrow \left( 
\begin{array}{c}
1-\mu _{*}x_{1}^{2} \\ 
1-\mu _{*}x_{2}^{2}
\end{array}
\right) 
\]
\[
f_{2}:\left( 
\begin{array}{c}
x_{1} \\ 
x_{2}
\end{array}
\right) \rightarrow \left( 
\begin{array}{c}
(1-c)x_{1}+cx_{2} \\ 
cx_{1}+(1-c)x_{2}
\end{array}
\right) 
\]
By the chain rule the Jacobian is 
\begin{equation}
J_{p}=\prod_{k=1}^{p}\left\{ \left( 
\begin{array}{cc}
-2\mu _{*}x_{1}(k) & 0 \\ 
0 & -2\mu _{*}x_{2}(k)
\end{array}
\right) \left( 
\begin{array}{cc}
1-c & c \\ 
c & 1-c
\end{array}
\right) \right\}  \label{5}
\end{equation}
with determinant 
\begin{equation}
\det J_{p}=(1-2c)^{p}(-2\mu _{*})^{2p}\prod_{k=1}^{p}x_{1}(k)x_{2}(k)
\label{2.4}
\end{equation}
Because of the permanence of the periodic orbits, for small $c,$ in the
neighborhood of the original coordinates (those for $c=0$), the product of
the last two factors in (\ref{2.4}) is uniformly bounded for all $p$. Then
for all sufficiently large $p$, $\left| \det J_{p}\right| <1$. The question
is how this overall contraction is distributed among the two eigenvalues of $%
J_{p}$.

To discuss the nature of the eigenvalues we may use a first order
approximation in $c$. For a periodic orbit of period $p=2^{n}$ we define 
\begin{equation}
X_{l,q}^{(i)}=\left\{ 
\begin{array}{lc}
(-2\mu _{*})^{q-l+1}\prod_{k=l}^{q}x_{i}(k) & \textnormal{if }q\geq l \\ 
1 & \textnormal{if }l>q
\end{array}
\right\}  \label{6}
\end{equation}
For small $c$ consider the linear approximation to the Jacobian $J_{p}$%
\[
\left( 
\begin{array}{cc}
(1-pc)X_{1,p}^{(1)} & c\sum_{k=1}^{p}X_{1,k}^{(1)}X_{k+1,p}^{(2)} \\ 
c\sum_{k=1}^{p}X_{1,k}^{(2)}X_{k+1,p}^{(1)} & (1-pc)X_{1,p}^{(2)}
\end{array}
\right) 
\]
The eigenvalues are 
\begin{equation}
\begin{array}{ccl}
\lambda _{\pm } & = & \frac{1}{2}(1-pc)\left(
X_{1,p}^{(1)}+X_{1,p}^{(2)}\right) \\ 
&  & \pm \frac{1}{2}\sqrt{(1-pc)^{2}\left(
X_{1,p}^{(1)}-X_{1,p}^{(2)}\right)
^{2}+4c^{2}\sum_{k=1}^{p}X_{1,k}^{(1)}X_{k+1,p}^{(2)}\sum_{k^{^{\prime
}}=1}^{p}X_{1,k^{^{\prime }}}^{(2)}X_{k^{^{\prime }}+1,p}^{(1)}}
\end{array}
\label{7}
\end{equation}
If the periodic orbit runs with the two coordinates $x_{1}$ and $x_{2}$
synchronized then 
\[
\begin{array}{ccc}
\lambda _{+} & = & X_{1,p} \\ 
\lambda _{-} & = & \left( 1-2pc\right) X_{1,p}
\end{array}
\]
and the orbit being unstable for $c=0$ it remains unstable for $c\neq 0$.
However if the two coordinates are out of phase by $\frac{p}{2}$ steps the
radical in Eq.(\ref{7}) is 
\[
\sqrt{\prod_{k=1}^{p}x(k)}\left| \sum_{i=1}^{p}x(i)x(i+1)\cdots x(i+\frac{p}{%
2}-1)\right| 
\]
The existence of a superstable orbit for all periods $p=2^{n}$ implies that
at $\mu =\mu _{*}$ the product $\prod_{k=1}^{p}x(k)$ has an odd number of
negative-valued coordinates. Therefore the two eigenvalues are complex
conjugate and, for small $c$, the contraction implicit in (\ref{2.4}) is
equally distributed by the two eigenvalues. Therefore for sufficiently large 
$N$ all orbits of this type with $p>2^{N}$ become stable periodic orbits.
$\Box$ 

The conclusion should not depend on the use, in the last step, of a linear
approximation (in $c$) because the sign of the term under the radical is a
consequence of the geometric nature of the orbits at the Feigenbaum
accumulation point. Then at sufficiently small $c$ the conclusions of the
linear analysis hold.

The attracting periodic orbits of the coupled system being associated to the
unstable periodic orbits of the Feigenbaum cascade, the basins of attraction
will be controlled by neighborhoods of these orbits in each coordinate.
Therefore a checkerboard-type structure is expected for the basins of
attraction.

Fig.1 shows the structure of the basins of attraction of the orbits for $%
c=4\times 10^{-4}$. A mesh of $4000\times 4000$, each point being an initial
condition for the two units. The coupled system is allowed to evolve for $%
4\times 10^{4}$ time steps and then the sequence is analyzed and its period
determined. Periods are detected up to $p_{\max }=2^{12}$. The basins,
labelled by the period of the orbits, are displayed using the color map in
Table I. The detected orbits have all periods from $2^{5}$ to $p_{\max }$
but some of the basins are so narrow that they can hardly be seen in the
figure.

\[
\textnormal{%
\begin{tabular}{|l|l||l|l|}
\hline\hline
\multicolumn{1}{||l|}{Period} & \multicolumn{1}{|l||}{Colour} & Period & 
\multicolumn{1}{|l||}{Colour} \\ \hline
$2^{4}$ & lightred & $2^{9}$ & cyan \\ \hline
$2^{5}$ & red & $2^{10}$ & lightblue \\ \hline
$2^{6}$ & yellow & $2^{11}$ & blue \\ \hline
$2^{7}$ & lightgreen & $>2^{12}$ & brown \\ \hline
$2^{8}$ & green & $others$ & darkgray \\ \hline
\end{tabular}
} 
\]

\begin{center}
\textnormal{Table I}
\end{center}

The behavior described above holds only for very small couplings.
Numerically one finds that, as the coupling strength increases, the number
of different stable periodic orbits decreases. A similar behavior occurs for
networks with many units where, for large couplings, global synchronization
or synchronization in clusters seems to be the rule as already been observed
in the past by many authors that have studied coupled map lattices.

\section{Feigenbaum networks with many units}

\subsection{Mean-field analysis}

For the calculations below it is convenient to use as variable the net input
to the units, $y_i=$ $\sum_jW_{ij}x_j$. Then Eq.(\ref{1.1}) becomes, 
\begin{equation}  \label{3.1}
y_i(t+1)=1-\mu _{*}\sum_{j=1}^NW_{ij}y_j^2(t)
\end{equation}

For practical purposes some restrictions have to be put on the range of
values that the connection strengths may take. For information processing
(pattern storage and pattern recognition) it is important to preserve, as
much as possible, the dynamical diversity of the system. That means, for
example, that a state with all the units synchronized is undesirable insofar
as the effective number of degrees of freedom is drastically reduced. From 
\begin{equation}
\delta y_{i}(t+1)=-2\mu _{*}y(t)(W_{ii}\delta y_{i}(t)+\sum_{j\neq
i}W_{ij}\delta y_{j}(t))  \label{3.2}
\end{equation}
one sees that instability of the fully synchronized state implies $\left|
2\mu _{*}y(t)W_{ii}\right| >1$. Therefore, the interesting case is when the
off-diagonal connections are sufficiently small to insure that 
\begin{equation}
W_{ii}>\frac{1}{\mu _{*}}  \label{3.3}
\end{equation}

For large $N$, provided there is no large scale synchronization effect, a
mean-field analysis might be appropriate, at least to obtain qualitative
estimates on the behavior of the network. For the unit $i$ the average value 
$<1-\mu _{*}\sum_{j\neq i}W_{ij}y_{j}^{2}(t)>$ acts like a constant and the
mean-field dynamics is 
\begin{equation}
z_{i}(t+1)=1-\mu _{i,eff}\bigskip \ z_{i}^{2}(t)  \label{3.4}
\end{equation}
where 
\begin{equation}
z_{i}=\frac{y_{i}}{<1-\mu _{*}\sum\limits_{j\neq i}W_{ij}y_{j}^{2}>}
\label{3.5}
\end{equation}
and 
\begin{equation}
\mu _{i,eff}=\mu _{*}W_{ii}<1-\mu _{*}\sum\limits_{j\neq i}W_{ij}y_{j}^{2}>
\label{3.6}
\end{equation}
$\mu _{i,eff}$ is the effective parameter for the mean-field dynamics of
unit $i$. From (\ref{3.3}) and (\ref{3.6}) it follows $\mu _{*}W_{ii}>\mu
_{i,eff}>\mu _{*}W_{ii}(2-\mu _{*})$. The conclusion is that the effective
mean-field dynamics always corresponds to a parameter value below the
Feigenbaum accumulation point, therefore, one expects the interaction to
stabilize the dynamics of each unit in one of the $2^{n}$- periodic orbits.
On the other hand to keep the dynamics inside an interesting region we
require $\mu _{i,eff}>\mu _{2}=1.3681$, the period-2 bifurcation point. With
the estimate $<y^{2}>=\frac{1}{3}$ one obtains 
\begin{equation}
\mu _{*}W_{ii}(1-\frac{\mu _{*}}{3}(1-W_{ii}))>\mu _{2}  \label{3.7}
\end{equation}
which, together with (\ref{3.3}), defines the interesting range of
parameters for $W_{ii}=1-\sum\limits_{j\neq i}W_{ij}$.

Except for direct numerical simulations and mean-field analysis, no other
efficient techniques are available for the study of coupled maps with many
units. However, as remarked by several authors \cite{Kaneko} \cite{Ershov},
the mean-field is in general a problematic assumption. In the numerical
simulations which we performed, to demonstrate the potential applications of
Feigenbaum networks as signal processors, we have indeed found that the
mean-field analysis holds only for a small range of (small) couplings. It
was in this narrow mean-field window that the numerical experiments were
carried out.

\subsection{Feigenbaum networks as signal processors}

Recent studies of biological systems suggest that complex and chaotic
dynamics plays an important role in biological information processing\cite
{ChaosBrain}. In the biological models the chaotic units are large
interacting aggregates of neurons not the individual neurons themselves. It
is therefore these aggregates that should be identified with the chaotic
units in the mathematical models.

In particular the evidence from the study of the mammalian olfactory system%
\cite{Freeman1} \cite{Skarda} \cite{Freeman2} \cite{Freeman3} \cite{Yao1} 
\cite{Yao2} is very interesting. The researchers that studied this system
propose a strange attractor with multiple wings, each one being accessed
depending on the initial conditions imposed by the external stimulus.
However with one invariant measure associated to a unique attractor (no
matter how many wings it possesses) we would expect the system to have a
non-negligible probability to explore all regions, leading to unreliable
pattern identification. It is probably better to consider that, to each
external stimulus, corresponds a different invariant measure, a scheme that
has recently been illustrated using Bernoulli units\cite{Dente}. Whatever
the exact dynamical mechanism, schemes of this type where, from a
low-intensity chaotic basal state, the system is excited into a different
chaotic attractor, have a very high storage capacity due to the diversity of
all possible dynamics and, on the other hand, display a very fast response
time. The recognition time in the olfactory system (in the sense of the time
needed to switch between two space-coherent patterns) being of the order of
magnitude of the response time of individual neurons, a chaos-based
mechanism is much more likely than the paradigm of attractor neural networks%
\cite{Amit}. A recognition time of the order of magnitude of the cycle time
is incompatible with the fixed point paradigm, but it is easy to understand
in a chaos-based scheme.

The Feigenbaum attractor is not chaotic, in the Lyapunov exponent sense.
Nevertheless, as an accumulation point of period-doubling bifurcations it
accesses an infinite number of different periodic orbits and, in principle,
is capable of processing a very large amount of information. As we have
seen, a system of many such units, interacting through synaptic-type
connections, is driven into periodic behavior by the coupling. The nature of
the orbits reflects the pattern of synaptic connections and if these are
obtained by a learning mechanism, through exposure to some external signal,
the system might act as a signal identifier. If, in addition, the external
signal is made to modulate some of the inputs to the individual units,
mechanisms of associative pattern recognition would be obtained.

Let, for example, the $W_{ij}$ connections be constructed from an input
signal $x_i$ by a correlation learning process 
\begin{equation}  \label{3.8}
\begin{array}{c}
W_{ij}\rightarrow W_{ij}^{^{\prime }}=(W_{ij}+\eta x_ix_j)e^{-\gamma }
\textnormal{
for }i\neq j \\ 
W_{ii}\rightarrow W_{ii}^{^{\prime }}=1-\sum_{j\neq i}W_{ij}
\end{array}
\end{equation}
The dynamical behavior of the network, at a particular time, will reflect
the learning history, that is, the data regularities, in the sense that $%
W_{ij}$ is being structured by the patterns that occur more frequently in
the data. The decay term $e^{-\gamma }$ insures that the off-diagonal terms
remain small and that the network structure is determined by the most
frequent recent patterns. Alternatively, instead of the decay term, we might
use a normalization method and the connection structure would depend on the
weighted effect of all the data.

In the operating mode described above the network acts as a {\it signal
identifier}. For example if the signal patterns are random, there is little
correlation established and all the units operate near the Feigenbaum point.
Alternatively the learning process may be stopped at a certain time and the
network then used as a {\it pattern recognizer}. In this latter mode,
whenever the pattern \{$x_i$\} appears, one makes the replacement 
\begin{equation}  \label{3.9}
\begin{array}{c}
W_{ij}\rightarrow W_{ij}^{^{\prime }}=W_{ij}x_ix_j \textnormal{ for }i\neq j \\ 
W_{ii}\rightarrow W_{ii}^{^{\prime }}=1-\sum_{j\neq i}W_{ij}
\end{array}
\end{equation}
Therefore if $W_{ij}$ was $\neq 0$ but either $x_i$ or $x_j$ is $=0$ then $%
W_{ij}^{^{\prime }}=0$. That is, the correlation between node $i$ and $j$
disappears and the effect of this connection on the lowering of the periods
vanishes.

If both $x_{i}$ and $x_{j}$ are one, then $W_{ij}^{^{\prime }}=W_{ij}$ and
the effect of this connection persists. Suppose however that for all the $%
W_{ij}$'s different from zero either $x_{i}$ or $x_{j}$ are equal to zero.
Then the correlations are totally destroyed and the network comes back to
the uncorrelated (nonperiodic behavior). This case is what is called a {\it %
novelty filter}. Conversely, by displaying periodic behavior, the network 
{\it recognizes} the patterns that are similar to those that, in the
learning stage, determined its connection structure. Recognition and {\it %
association} of similar patterns is then performed.

A numerical simulation was made for a network of $100$ units. During a
learning phase a single binary pattern $\{x_{i}\}$ was presented many times
to the network, the network connections evolving according to the
correlation learning method (Eq.(\ref{3.8})) and a normalization
prescription.

After the learning phase, the connections $W_{ij}$ are considered to be
frozen and we have tested the reaction of the network to several other
patterns. In this recall phase, to present a pattern to the network means to
make the replacements of Eqs.(\ref{3.9}) and to let the system evolve from a
random initial condition. As the system might be sensible to the initial
conditions, we choose the same initial conditions for all patterns. The
results are shown in Fig.2. Each pattern leaves a unique signature given by
the set of periods to which each unit converges. The periods in each unit
are measured after a transient delay of $2\times 10^{4}$ time steps.

In Fig.2 the first pattern that is presented to the network is the one that
was memorized. The others are chosen in the following way:

- the second is the complementary binary pattern to the one that was
memorized. This pattern being non-correlated to the learned pattern, the
network units become uncoupled and only very high periods are seen.

- for the remaining patterns some bits are swapped (from $0$ to $1$) in the
complementary pattern, forming new signals which are weakly correlated with
the memorized one. The bit swapping structure is listed in Table II. The
signature of the correlations is seen from the low periods that appear in
some of the network units (Fig.2). The color map is the same as used in
Sect. 2.

\begin{center}
\[
\begin{tabular}{|l||l|}
\hline\hline
\multicolumn{1}{||l||}{pattern} & \multicolumn{1}{||l||}{correlated bits} \\ 
\hline
memorized & all \\ \hline
$1$ & none \\ \hline
$2$ & $10,76$ \\ \hline
$3$ & $10,20,76,89$ \\ \hline
$4$ & $5,10,20,36,52,76,89,100$ \\ \hline
\end{tabular}
\]

\textnormal{Table II}
\end{center}

\section{Final remarks}

For its dynamical diversity and the selective stabilization effect of the
couplings, Feigenbaum networks, both with few and with many units, seem to
be a potentially interesting system for signal processing through complex
dynamics and to model self-organization.

As seen in Sect.2, the size of the basins of attraction of the sinks created
by the coupling is controlled by the geometry of the pre-existing unstable
periodic orbits and, as a consequence, are extremely small for large
periods. It might therefore be useful for practical applications to be able
to control the dynamics in a more efficient way. This implies departures
from the simple linear coupling scheme of Eq.(\ref{1.1}). For example we
might use 
\begin{equation}  \label{4.1}
x_i(t+1)=1-\mu _{*}\left\{ x_i(t)+\sum_{j\neq i}W_{ij}f(x_i(t),x_j(t))\left(
x_j(t)-x_i(t)\right) \right\} ^2
\end{equation}
with 
\[
f(x,y)=\theta (x)\theta (x-y)+\theta (-x)\theta (y-x) 
\]
or the smooth version 
\[
f(x,y)=\frac 12\left\{ 1+\tanh \left( \frac k2x\right) \tanh \left( \frac
k2(x-y)\right) \right\} 
\]
For small couplings this leads to a much larger shift to smaller values of
the $\mu _{i,eff}$ discussed in Sect.3.

Alternatively we might think of using other maps, which also possess an
infinite diversity of periodic behaviors. Consider for example a
piecewise-linear dynamics for the units ($x\rightarrow \beta x+\alpha $ (mod.%
$1)$). For $\beta >1$ the uncoupled dynamics is fully chaotic. However,
through interaction, it is possible to select and stabilize an infinite
number of periodic orbits. Consider a fully connected network with dynamics 
\begin{equation}  \label{4.3}
x_i(t+1)=\beta \left( \sum_jW_{ij}x_j(t)\right) +\alpha \bigskip\ \textnormal{ }%
\bigskip\ \textnormal{(mod. 1)}
\end{equation}
with $x_i$ in the interval $[0,1)$. For a large number of weakly correlated
units a mean-field analysis as in Sect.3 leads to 
\begin{equation}  \label{4.4}
x_i(t+1)=\beta _{i,eff}W_{ii}x_i(t)+\alpha _{i,eff}\bigskip\ \textnormal{ }%
\bigskip\ \textnormal{(mod. 1)}
\end{equation}
where 
\begin{equation}  \label{4.5}
\beta _{i,eff}=\beta W_{ii}
\end{equation}
\begin{equation}  \label{4.6}
\alpha _{i,eff}=\alpha +\beta \sum_{j\neq i}W_{ij}x_j
\end{equation}
If $\beta >1$ the individual uncoupled units have an invariant measure
absolutely continuous with respect to Lebesgue measure. However if $\beta
_{i,eff}<1$ the dynamics of Eq.(\ref{4.4}) has attracting periodic orbits of
all periods, depending on the values of the parameters $\alpha _{eff}$ and $%
\beta _{eff}$ . The point of the periodic orbit with the smallest coordinate
is obtained from\cite{Vilela} 
\begin{equation}  \label{4.7}
(1-\beta _{i,eff}^p)x_{\min }=\alpha _{i,eff}\frac{(1-\beta _{i,eff}^p)}{%
(1-\beta _{i,eff})}-P_p(\beta )
\end{equation}
where $p$ is the period of the orbit and $P_p(\beta )$ is a polynomial in $%
\beta $%
\begin{equation}  \label{4.8}
P_p(\beta )=1+\sum_{k=2}^{p-1}\left\{ \left[ \frac{kj}p\right] -\left[ \frac{%
(k-1)j}p\right] \right\} \beta ^{p-k}
\end{equation}
where $j$ (prime to $p$) is the jump number of the orbit and $\left[
n\right] $ denotes the integer part of $n$.

{\bf Figure captions}

Fig.1 - Color coded basin of attraction periods for two units and $c=4\times
10^{-4}$

Fig.2 - Color coded periods in a network of $100$ units reacting to several
patterns

\end{document}